\documentclass[aps,pra,reprint,showpacs,amsmath,amssymb]{revtex4-1}
\usepackage{graphicx}

\begin{document}

\title{Effects of depolarizing quantum channels on BB84 and SARG04 quantum cryptography protocols}

\author{Youn-Chang Jeong}\email{w31400@gmail.com}
\affiliation{Department of Physics, Pohang University of Science and Technology (POSTECH), Pohang, 790-784, Korea}

\author{Yong-Su Kim}
\affiliation{Department of Physics, Pohang University of Science and Technology (POSTECH), Pohang, 790-784, Korea}

\author{Yoon-Ho Kim}\email{yoonho72@gmail.com}
\affiliation{Department of Physics, Pohang University of Science and Technology (POSTECH), Pohang, 790-784, Korea}

\date{\today}

\begin{abstract}
We report experimental studies on the effect of the depolarizing
quantum channel on weak-pulse BB84 and SARG04 quantum cryptography.
The experimental results show that, in real world conditions in which
channel depolarization cannot be ignored, BB84 should perform better
than SARG04.
\end{abstract}

\pacs{03.67.Dd, 03.67.Hk, 42.79.Sz}

%\keywords{Quantum cryptography, BB84 protocol, SARG04 protocol,Quantum communication, depolarizing channel}
%Use show keys class option if keyword display desired

\maketitle

%%%%%%%%%%%%%%%%%%%%%%%%%%%%%%%%%%%%%%%%%%%%%%%%%%%%%%%%%%%%%%
%%%%%%%%%%%%%%%%%%%%%%%%%%%%%%%%%%%%%%%%%%%%%%%%%%%%%%%%%%%%%%

In quantum cryptography, the goal is to establish shared secret keys between two communicating parties, Alice and Bob \cite{Gisin2002}. One of the key components of the original BB84 quantum cryptography protocol is the single-photon source, but the lack of efficient single-photon sources resulted experimental implementations with weak pulses of light \cite{BB84,B92}. The weak pulse implementations are practical but is susceptible to the photon number splitting (PNS) attack \cite{B92, Huttner1995}. It, nevertheless, has been shown that secure keys can be extracted from the weak pulse implementations of the BB84 protocol if certain conditions are met \cite{brassard,lut}.

There also exist quantum cryptography protocols which are inherently more secure against the PNS attack when implemented with weak pulses. One is the decoy state method \cite{Hwang2003} and the other is the SARG04 protocol \cite{SARG04}. The decoy state method, while very robust against the PNS attack, requires additional physical resources which could introduce further vulnerabilities to the system \cite{decoy1,decoy2,decoy3}. The SARG04 protocol, on the other hand, offers more robustness against the PNS attack than the BB84 protocol implemented with weak pulses, while using the same hardware as that of the BB84 protocol \cite{multiphoton}.

One important problem in real-world quantum cryptography systems is the systems' behaviors under the presence of disturbances  in the quantum channel (e.g., the depolarization effect, the phase error, damping, etc.)  connecting Alice and Bob. In particular, it is of importance to compare the performances for the  BB84 and SARG04 quantum cryptography systems as they differ only in the classical key sifting procedures \cite{compare,compare2,lowerbound}. In this paper, we investigate experimentally how the depolarization effect in the quantum channel affects the secure key generation performances for the weak-pulse BB84 and SARG04 quantum cryptography systems.

%%%%%%%%%%%%%%%%%%%%%%%%%%%%%%%%%%%%%%%%%%%%%%%%

In the weak-pulse implementation of the BB84 and the SARG04 quantum cryptography protocols, the qubit $|\Psi\rangle$ is  encoded in the polarization state of a weak pulse of light and is sent to Bob via a free-space or a fiber quantum channel.  The effect of depolarization in the quantum channel on the qubit $|\Psi\rangle$ can be written as $\varepsilon[|\Psi\rangle]$ and is given by\cite{compare}
%%%%%%%%%%%%%%%%%%%%%%%%%%%%%%%%%%%
\begin{equation}\label{depolarizing}
\varepsilon [|\Psi \rangle ] = F |\Psi \rangle \langle\Psi| + D
|\Psi^{\perp} \rangle \langle\Psi^{\perp}|,
\end{equation}
%%%%%%%%%%%%%%%%%%%%%%%%%%%%%%%%%%%%%
where $|\Psi^\perp\rangle$ is the state orthogonal to $|\Psi\rangle$, $F$ is the fidelity of the quantum channel, and $D$ quantifies the disturbance in the quantum channel. The values $F$ and $D$ depend on the quantum channel visibility $V$ as $F =(1 + V)/2$ and $D = (1 - V)/2$, respectively \cite{compare}. Note that $F+D=1$.

Let us now consider how the channel depolarization will affect the quantum bit error rate (QBER, $Q$), which is the ratio of the number of error bits in the sifted key to the total number of sifted keys. In the BB84 protocol, a sifted key bit is generated when Alice and Bob happen to choose the same basis and, thus,  the error bit in the sifted key is dependent on the disturbance $D$ in the quantum channel. Since the probabilities for generating the error bit and the correct bit in the BB84 protocol are given as $p_e=D$ and $p_c=F$, respectively, QBER can be written as\cite{compare},
%%%%%%%%%%%%%%%%%%%%%%%%%%%%%%%%
\begin{equation}\label{BB84QBER}
Q^{BB84} = \frac{p_{e}}{p_{c} + p_{e}} = \frac{D}{F + D}
= \frac{1 - V}{2}.
\end{equation}
%%%%%%%%%%%%%%%%%%%%%%%%%%%%%%%

In the SARG04 protocol, the situation is a bit different. Consider, for example, Alice sends $|V\rangle$
and announces publicly $S_{0} \equiv \{|V\rangle, |45^\circ\rangle\}$. On Bob's side, out of four possible outcomes, only the $|-45^\circ\rangle$ outcome is a conclusive one \cite{SARG04}. When depolarization in the quantum channel is considered,  $\varepsilon [|V\rangle]=F|V\rangle\langle V|+D|H\rangle\langle
H|=\frac{1}{2}|+45^\circ\rangle\langle
+45^\circ|+\frac{1}{2}|-45^\circ\rangle\langle
-45^\circ|+\frac{F-D}{2}|45^\circ\rangle\langle
-45^\circ|+\frac{F-D}{2}|-45^\circ\rangle\langle 45^\circ|$. Thus, the probability of the correct bit is $p_c=1/2$ and is independent of the channel visibility $V$. The error bit probability, however, depends on the disturbance factor $D$ in the quantum channel, $p_e=D$. QBER is thus given as \cite{compare}
%%%%%%%%%%%%%%%%%%%%%%%%%%%%%%%%%%%
\begin{equation}\label{SARG04QBER}
Q^{SARG04} = \frac{D}{\frac{1}{2} + D} = \frac{1 - V}{2 - V}.
\end{equation}
%%%%%%%%%%%%%%%%%%%%%%%%%%%%%%%%%%

The total sifted key rates $R_\mu$ for the weak-pulse BB84 and the SARG04 protocols are given as \cite{lowerbound}
%%%%%%%%%%%%%%%%%%%%%%%%%%%%%%%%%%%
\begin{eqnarray}
R^{BB84}_{\mu} & = & \frac{1}{2}(1-\overline{p}_d^2 e^{-\mu \eta}),
\label{BB84Gain}
\\
R^{SARG04}_{\mu} & = & \frac{1}{2} ( 1 +
\frac{\overline{p}_d}{2}e^{-\mu F \eta} -
\frac{\overline{p}_d}{2}e^{-\mu D \eta}
- \overline{p}^2_d e^{-\mu \eta}),\label{SARG04Gain}
\end{eqnarray}
%%%%%%%%%%%%%%%%%%%%%%%%%%%%%%%%%%%
for the average photon number per pulse $\mu$ and for a quantum channel characterized with the transmission factor $t = 10^{-\alpha l /10}$ where $\alpha$ is the attenuation coefficient (dB/km) and $l$ is the length of the channel. Also, $\overline{p}_d = 1 - p_d$ where $p_d$ is the dark count probability and $\eta = \eta_{d} t $ where $\eta_{d}$ is Bob's detection efficiency. Note that $R_\mu =\sum_n R_n = \sum_n Y_n e^{-\mu} \mu^n/n!$ where $Y_n$ is the probability for Bob to have a conclusive result for the n-photon pulse sent from Alice. It is interesting to point out that the sifting rate for BB84 is independent of the channel visibility $V$ while that of SARG04 varies with $V$. This is due to the fact that, for BB84, $p_e+p_c=F+D=1$, but for SARG04, $p_{e}+p_{c}=1/2+D$.

%%%%%%%%%%%%%%%%%%
\begin{figure}[t]
\includegraphics[width=3.4in]{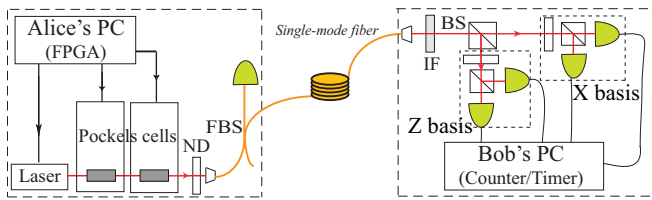}
\caption{Schematic of experimental setup. See text for details.}\label{setup}
\end{figure}
%%%%%%%%%%%%%%%%%%%

The secure key rates (the lower bound) in the presence of depolarization in the quantum channel can then be determined with the total sifting rates $R_\mu$ and the average QBER $Q_\mu \equiv \sum_n  Q_n R_n/R_\mu$ ($Q_n$ denotes the QBER for the n-photon pulse) using experimental obtained parameters \cite{lowerbound}.

To experimentally test the effects of depolarization in the quantum channel on the performance of weak-pulse BB84 and SARG04 quantum cryptography protocols, we have built a fiber-based quantum cryptography system schematically shown in Fig.~\ref{setup}. Alice's setup consists of a diode laser which emits a train of 3 ns long 780 nm laser pulses at 1 MHz repetition rate, two Pockels cells for polarization encoding, and FPGA-based electronics for controlling and storing the data. Polarization-encoded laser pulses are then attenuated with a neutral density (ND) filter set and coupled into a fiber beamsplitter (FBS). A single-photon detector connected to one end of FBS is used to measure the average photon number per pulse so that $\mu<1$ and the other end of FBS is connected to the quantum channel, a 1.27 km long single-mode fiber spool. The attenuation coefficient of the fiber at 780 nm is $\alpha = 3$ dB/km and the initial channel visibility of the fiber quantum channel is measured to be $V_i = 0.954 \pm 0.002$. Bob's setup consists of a 3nm bandpass filter (IF), a 50/50 beam splitter for random selecting between Z-basis ($|H\rangle$ and $|V\rangle$) and X-basis ($|45^\circ\rangle$ and $|-45^\circ\rangle$), four Si single-photon detectors and a counter/timer card for data storage. Bob's detection efficiency (including optics and the detector) and the dark count probability are measured to be $\eta_{d} = 0.4$ and $p_{d} = 3.3 \pm 0.6\times 10^{-5}$, respectively.

Since depolarization in the quantum channel is related to the channel visibility as $D=(1-V)/2$, it is necessary to vary the channel visibility $V$ to observe the effect of depolarizing quantum channels on BB84 and SARG04 protocols. Rather than actually introducing additional fiber-based polarization controllers to the setup to decrease the channel visibility from the initial value $V_i = 0.954$, we use the effective method described as follows. Note that the effect of quantum channel depolarization is to flip the qubit $|\Psi\rangle$ to the orthogonal one $|\Psi^{\perp}\rangle$ with the probability $D$. Thus, to effectively implement channel depolarization, we simply flip the stored bit information at Alice with probability $D$ determined by  the desired channel visibility value. For example, to achieve the effect of channel visibility of 0.9, 5\% of Alice's stored bits are randomly selected and flipped. Note that, in this case, since the initial channel visibility is 0.954, we are in fact achieving the channel visibility of 0.859.

%%%%%%%%%%%%%%%%%%%%%%
\begin{figure}[t]
\includegraphics[width=3in]{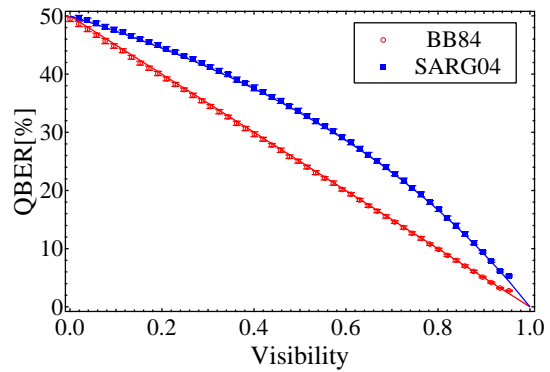}
\caption{QBER vs. channel visibility. The solid lines are due to eq.~(\ref{BB84QBER}) and eq.~(\ref{SARG04QBER}). Experiments are done for $\mu= 0.189 \pm 0.001$}\label{QBER}
\end{figure}
%%%%%%%%%%%%%%%%%%%%%

% Fig. 2: QBER vs channel visibility

Figure \ref{QBER} shows QBER as a function of channel visibility for the BB84 and the SARG04 protocols in the case of average photon number per pulse $\mu = 0.189$. The experimental results shown in Fig.~\ref{QBER} agree well with eq.~(\ref{BB84QBER}) and eq.~(\ref{SARG04QBER}). We have repeated the QBER vs. channel visibility measurement for various values of the average photon number per pulse, $\mu \approx 0.03 \sim 0.30$, and they show the same results as in Fig.~\ref{QBER}. The SARG04 protocol, therefore, is shown to be more sensitive than the BB84 protocol for the depolarization effect in the quantum channel.

% Fig. 3: Sifted key vs V

\begin{figure}[t]
\includegraphics[width=3in]{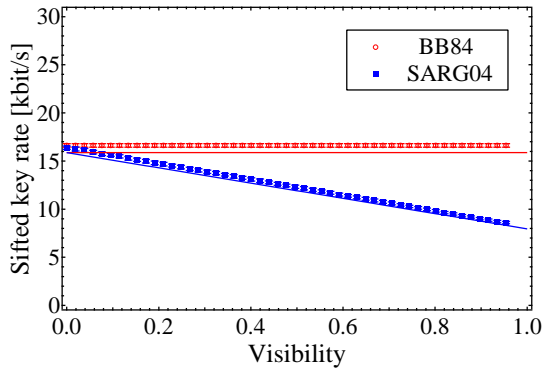}
\caption{Sifted key generating rate vs. quantum channel visibility.  The solid lines are due to eq.~(\ref{BB84Gain}) and eq.~(\ref{SARG04Gain}). Experiments are done for $\mu = 0.189 \pm 0.001$
}\label{Siftedkey}
\end{figure}

\begin{figure}[t]
\includegraphics[width=3in]{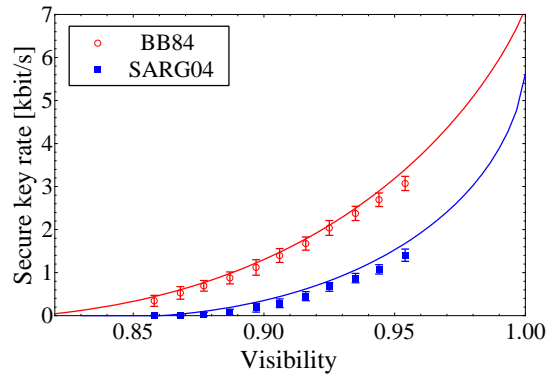}
\caption{Secure key generation rate (the lower bound) vs. channel visibility. The BB84 protocol is shown to generate more secure keys than the SARG04 protocol in the presence of depolarization in the quantum channel. The solid lines are theoretical plots due to the results in Ref.~\onlinecite{lowerbound} using the parameters of the experimental setup.
}\label{Securekey}
\end{figure}

We have also tested how the sifted key rates are affected by the depolarizing effect in the quantum channel. The experimental results shown in Fig.~\ref{Siftedkey} reveal that the sifted key rate for the BB84  is independent on the channel visibility, but the SARG04 protocol exhibits channel visibility dependent sifted key rates. It is also worth while to mention that, as evident from Fig.~\ref{Siftedkey}, the BB84 protocol always generates more sifted keys than the SARG04 protocol. Note that Fig.~\ref{Siftedkey} shows small discrepancies between the experimental results and the theoretical results due to eq.~(\ref{BB84Gain}) and eq.~(\ref{SARG04Gain}). These discrepancies suggest that there are slight systematic errors in estimating the average photon number per pulse $\mu$, the detection efficiencies, and the dark counts. These errors, however, are small enough so that they do not affect the secure key generation rates (the lower bound) significantly, see Fig.~\ref{Securekey}.

%%%%%%%%%%%%%%%%%%%%%%%
%%%%%%%%%%%%%%%%%%%%%%%

%visibility vs overall error bit rate

Although the sifted key rates for the BB84 and the SARG04 protocols behave differently in the presence of depolarization in the quantum channel, the overall error rates $R_{\mu} Q_{\mu}$ are shown to be the same for a given channel visibility  \cite{lowerbound},
%%%%%%%%%%%%%%%%%%%%%%%%%%%%%%%%%%%%%
\begin{eqnarray}\label{GainQBER}
\lefteqn{R^{BB84}_{\mu} Q^{BB84}_{\mu} =  R^{SARG04}_{\mu}
Q^{SARG04}_{\mu}} \nonumber \\
&  & = \frac{1}{4} ( 1 + \overline{p}_d e^{-\mu F \eta} -
\overline{p}_d e^{-\mu D \eta} - \overline{p}^2_d e^{-\mu \eta} ).
\end{eqnarray}
%%%%%%%%%%%%%%%%%%%%%%%%%%%%%%%%%%%%
We have calculated the overall error bit rates $R_{\mu} Q_{\mu}$ for a variety of values of the quantum channel visibility using the data shown in Fig.~\ref{QBER} and Fig.~\ref{Siftedkey} and found that the above relation holds. The overall error rate is shown to decrease linearly as the visibility of the quantum channel is increased.

% Fig. 4: Secure key vs V

Finally, we study the secure key generation rates (the lower bound) for the BB84 and the SARG04 protocols in the presence of depolarization in the quantum channel. As mentioned earlier, the secure key rates are calculated from the experimentally obtained QBER, sifted key rates, and parameters of the experimental setup \cite{lowerbound}. Figure \ref{Securekey} shows the the secure key generation rates (the lower bound) for the two protocols as a function of the channel visibility. Note that each data point in Fig.~\ref{Securekey} corresponds the maximum value of the secure key rates (the lower bound) for the range of values of the average photon number per pulse, $0.03 \leq \mu  \leq  0.30$, for a given channel visibility. In other words, each data point in Fig.~\ref{Securekey} is obtained with the optimum value of $\mu$. We also note that the optimum value of $\mu$ is shown to decrease monotonically (for BB84, from $\mu=0.189$ to $\mu=0.060$; for SARG04, from  $\mu=0.189$ to $\mu=0.033$) as the quantum channel visibility is reduced.

%%%%%%%%%%%%%%%%%%%%%%%%
%%%%%%%%%%%%%%%%%%%%%%

The experimental results in Fig.~\ref{Securekey} show that, although the SARG04 protocol has been known to be more robust against the PNS attack than the BB84 protocol, the latter performs better in real-world conditions, i.e., limited detection efficiency, non-zero dark count probability, and, most importantly, the quantum channel subject to depolarization effects.

In conclusion, we have demonstrated experimentally the effect of depolarization in the quantum channel  on weak-pulse BB84 and SARG04 quantum cryptography protocols. The experimental results show that the SARG04 protocol is more susceptible to the depolarization effect in the quantum channel than the BB84 protocol. In real-world quantum cryptography applications in which the quantum channel cannot be assumed to be perfect, therefore,  the BB84 protocol appears to provide better performance than the SARG04 protocol.

%%%%%%%%% Ack %%%%%%%%%%%
This work was supported, in part, by the National Research Foundation of Korea (R01-2006-000-10354-0 and  2009-0070668) and POSTECH BSRI Fund. YSK acknowledges the support of the Korea Research Foundation (KRF-2007-511-C00004).

%%%%%%%%%%%%%%%%%%%%%%%%%%%%%%%%%%%%%%

\end{document}